# Quasi-two-dimensional metallic hydrogen inside di-phosphide at high pressure

Nikolay Degtyarenko, Evgeny Mazur

*National Research Nuclear University "MEPHI", Kashirskoe sh.31, Moscow 115409, Russia*

**Abstract**

The method of mathematical modeling was used for the calculation of the structural, electronic, phonon, and other characteristics of various normal phases of phosphorus hydrides with stoichiometry $PH_k$. It was shown that the di-phosphine may form 2D lattice of the metallic hydrogen, stabilized by phosphorus atoms under high hydrostatic compressive pressure. The resulting structure with the elements of H-P-H is a locally stable one in the phonon spectrum (or metastable). The properties of di-phosphine with the properties of similar structures such as thesulphur hydrides were compared.

## 1. Introduction

One of the areas of search for new superconducting materials with high critical temperatures is the study of compounds with high hydrogen content. In [1] it was discovered, that the transition to the superconducting state of the metallic phase of the hydrogen sulfide under pressure 170GPatakes place at the temperature T=203K. The idea is based on the theoretical works on the study of the metallic hydrogen properties, which predicted high critical temperature Tc ~ 200 ÷ 400K [2]. In the works [3, 4] it was shown that such a phase of hydrogen may be metastable. After publication of the results [1] the interests in searching for compounds that can go into a superconducting state at higher temperatures have increased significantly. In particular, the preliminary experimental results on the compression of phosphide $PH_3$ and its superconductivity at 100K [5] should be mentioned. In [1] the fact is considered that at the high pressure the hydrogen forms a hydride $SH_n$ (n≤3) [6]through dissociation of the sulfur compounds. The properties of the $SH_3$in the hydrogen sulfide can lead to the possible explanation ofthe high-temperature superconductivity existence. In [7-9] the $SH_2$ phase with the symmetry I4/MMM was found along with the compounds $SH_3$, for which the electron and phonon spectra and the density of electron and phonon states in the pressure range 100÷225GPa were calculated.

In this paper we present the calculations of the electronic and phonon properties of the di-phosphine H2P-PH2 system compressed withup to 250 GPA hydrostatic pressures. Di-phosphine under normal conditions is a colorless liquid with the structure of molecules presented in Fig.1A [10].

## 2. Method.

The calculations of the structural, electronic and phonon characteristics of the molecules and the periodic crystal structure of $PH_k$ in the pressure range 0÷ 250GPa were performed using ab-initio method. The DFT approximation with the plane wave basis with the correlation functional GGA – PBE pseudopotential which preserves the norm was used. The unit cell of the structure used to calculate the electronic properties contains an element of H2P-PH2. All the calculations were performed in the spin-polarized approximation in order to compare the calculated energy values of different phases in an adequate manner.

## 3. Results.

At the 25GPa pressure the molecular di-phosphine $P_2H_4$phase becomes metallic because of the overlap of electron shells of the molecules. The compressionof the combination of such molecules in thepressure range 25 ÷ 250GPa leads to the formation of metallic crystalline phase. The main structural element of this phase H2P-PH2 acquires straight H-P-H bond directed at an angle of 180° (Fig. 1.c). In the calculations of the first cycle, the pressure for the molecular phase was assumed to be the source of the initial cell symmetry with the P2H4 molecule havingthe symmetry P1. The resulting crystal structure at the pressure of 25÷125 GPA acquires the symmetry of the unit cell of I4/MMM. The structure formed in this range of pressures with a finalI4/MMM crystal symmetry (Fig.1.e) as a result of the structural modification of the unit cell in the studied phase with the stoichiometry $P_2H_4$. At the next stage of the calculations the cell with the symmetry I4/MMM was used as a starting one (second cycle). Crystal structure with this unit cell symmetry was studied in the pressure range 25÷250GPa. The crystal forms a set of parallel planes of atoms with the full concentration of all hydrogen atoms in these planes (Fig.1.e) in the entire pressure range 25÷250GPa as a result of structural modification of the unit cell in the studied phase of $P_2H_4$ with the symmetry I4/MMM. The hydrogen atoms form a square lattice with a period depending on the pressure in these planes. As a result, the electronic properties of the system acquire quasi-two-dimensional character. The phosphorus atoms form theirown planes. Each phosphorus atom has four atoms Pin the first coordination sphere located in the plane, with two hydrogen atoms above and below the plane. The calculated unit cell volume at the pressure P=125 GPA amounted to V=13.54$A^3$, which gives the total electron density value of n=$10^{24}$ elect./$cm^3$ (14 electrons per cell, or about 1 elect./$A^3$). The effect of the redistribution of the electron density between atoms with increasing pressure, which is determined by the conditional Mullikencharges of phosphorus $Z_M(P)$and hydrogen(H), should be considered as interesting one. Mullikencharges of atoms of phosphorus $Z_M(P)$ are positive with the increasing in magnitude character. The hydrogen increases the negative charge $Z_M(H)$. It follows from the results of calculations that the pressure P "pushes away" the electron density from the areas of the loops of the electron density around the phosphorus atoms in the hydrogen atoms. Also the results of the calculations show the distance **h** change between two approaching planes, in which hydrogen atoms forming quasi-two-dimensional square lattice with period a are concentrated, with increasing pressure P. The total concentration of hydrogen atoms within the planes occurs when the pressure approaches to the value P=150GPa. The Fig.2 shows the properties of the electron subsystem for the pressure P= 175GPa. Fig.2 shows that the Fermi level is located at the electronic DOS peak value. Fig.2 also shows that at the pressure P= 175GPa (Fig.2) at least one of the zones has dropped below the Fermi level. The estimation of effective masses of carriers near the Fermi level gives a value of about 0.1$m_{e0}$. At the pressure P=175GPA the Fermi level crosses three subzones (Fig.2).

There are no imaginary frequencies in the pressure range 50÷250GPa (Fig.3) in the phonon spectrum of the crystal $P_2H_4$ – I4/MMM phase. That means that the phase $P_2H_4$ – I4/MMM is a stable one in this pressure range. Maximum frequencies lie in the energy region of the order of 265meV. Note that the optical modes of oscillation corresponding to the displacement of hydrogen atoms (Fig.3)correspond not only to small q (point G), but also to all values of the wave vector q on segments connecting the symmetry points of the Brillouin zone (Fig.1.c). This range becomes more rigid with increasing pressure.

Fig. 4 shows graphs of the enthalpy ΔH with increasing pressure P for the most stable crystalline $SH_2$ and $SH_3$sulfide, di-phosphide $P_2H_4$ and phosphine $PH_3$ phases. Fig. 4 shows that compounds of the type $XH_2$ are metastable in concern with the phase transition in phase $XH_3$, for which enthalpy has smaller values. Presumably, for such a phase transition in experiments [1] the heating of the contents of the diamond cells with laser radiation was used. If this heating is eliminatedwhen compressing di-phosphine $P_2H_4$, then the phase with the symmetry I4/MMM can be formed in a metastable state. Indeed, its frequency spectrum in the range 50÷250GPa does not contain imaginary frequencies. In addition, the range of pressures P, in which the structure with stoichiometry of phosphine $PH_3$ and the symmetry IMMM has no imaginary frequencies, is relatively small – 100 ÷150GPa.

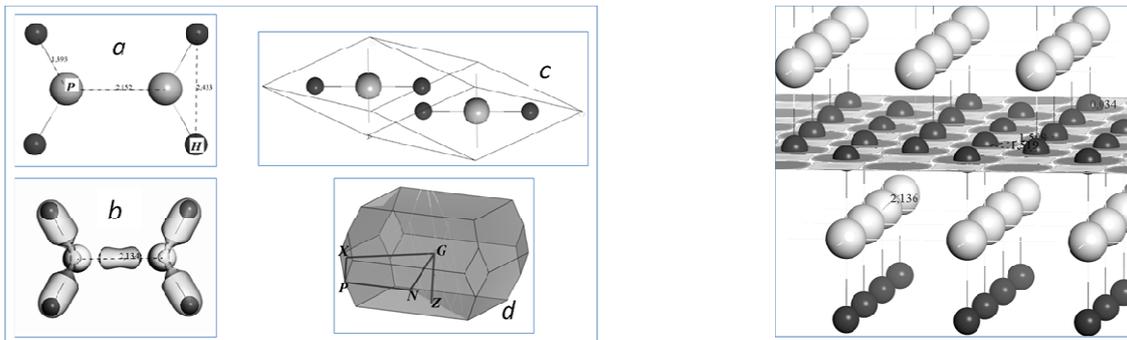

Fig.1. a – structure of di-phosphine under normal conditions; b – isosurface of the electron density for molecules of di-phosphine at the level of 0.6, indicating the covalent bonding in the molecule; c is a primitive cell with two elements of H-P-H at the pressure P=150GPa; d is the Brillouin zone corresponding to the primitive cell at the pressure P =150GPa; e is the structure formed from molecules of di-phosphine at the pressure P=150GPa; The two-dimensional distribution of electron density in the plane with the concentration of hydrogen atoms is shown.

4. Conclusions

The $P_2H_4$ di-phosphine molecular phase becomes metallic with pressure having values more than 25GPa because of the overlap of the electron shells of molecules. The crystal forms a set of parallel planes of atoms in the range of pressures P = 50÷250GPa with the full concentration of all hydrogen atoms in these planes as a result of the structural modification of the unit cell in the studied $P_2H_4$ phase with the final symmetry I4/MMM (Fig.1.e) with the formation of a square lattice of hydrogen atoms in this planes. As a result, the electronic properties of the system acquire quasi-two-dimensional metallic character. There are no imaginary frequencies in the phonon spectrum of the crystal di-phosphine $P_2H_4$ – I4/MMM phase in the pressure range 50÷ 250GPa.

$P_2H_4$ phase with the I4/MMM symmetry is stable (or metastable) in the pressure range 50÷250GPa. The maximum frequency lies in the 265meV value energy region. The range becomes more rigid with increasing pressure. The competing $PH_3$ phase crystal phosphine structure has a lower enthalpy at the given pressure P. The pressure range of its stability narrows the interval of pressure values to the following one: P=100÷150GPa.

Hydrostatic pressure not only forms a two-dimensional plane with the full concentration of hydrogen, but also "pushes away" the electron density to the many-electron atoms (sulfur or phosphorus) in these planes. Conventionally, this effect can be described as a "doping" of this plane.

Basing on the calculations of properties of the normal phase of di-phosphine we conclude that the optimal pressure range for the effect of superconductivity is the value about 150GPa. It seems to us that under hydrostatic compression the substance must show a transition to the superconducting state at a higher temperature than $PH_3$ [5] because it has the properties of two-dimensional metallic hydrogen in a wide range of pressures.

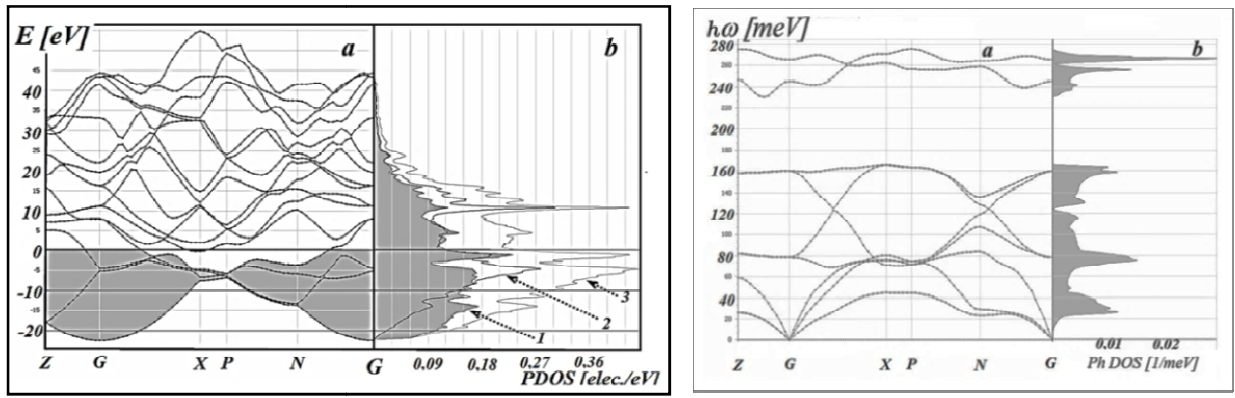

Fig.2. (left)  *a* – band structure of $P_2H_4$ electrons at the pressure P= 175GPA; *b* – PDOS, the partial density of the number of states:1 – s–electron States;2 – p- electron States;3 – the total (s,p) DOS.

Fig.3. (right) *a* – phonon dispersion curves $P_2H_4$ at the pressure P=175GPA; *b* – PhDOS – phonon $P_2H_4$ density of states, the maximum energy of the phonons at 175GPa – hω≈270meV.

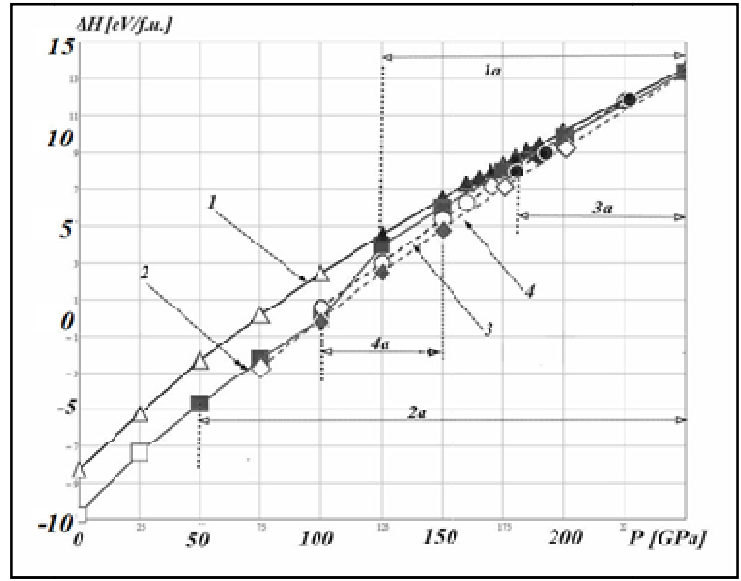

Fig.4. The enthalpy change ΔH of different phases with increasing pressure P: 1 – sulfide $SH_2$ (symmetry I4/MMM); 2 – di-phosphine $P_2H_4$ (symmetry I4/MMM); 3 – $SH_3$ (symmetry IM-3M); 4 – $PH_3$ (symmetry IMMM); the empty symbols correspond to the presence of imaginary frequencies, and filled symbols correspond to the absence of imaginary frequencies;
1a, 2a, 3a and 4a – the intervals of the pressure P, in which the structures have only real frequencies.

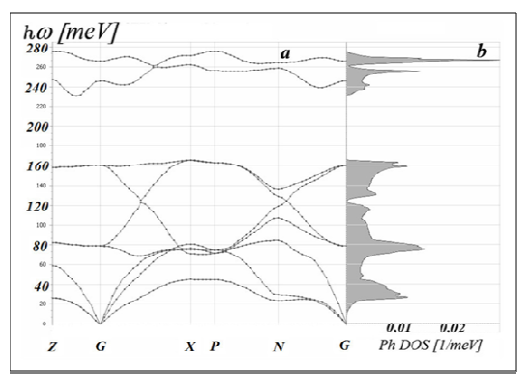